\documentclass[twocolumn,showpacs,preprintnumbers,amsmath,amssymb]{revtex4-1}

\usepackage{graphicx}
\usepackage{dcolumn}
\usepackage{bm}
\usepackage{subfigure}
\usepackage{color}

\graphicspath{{./figures/final/}}

\newcommand{\RAA}{\AA$^{-1}$}

\def\nba#1{}

\def\comment#1{}
\def\delete#1{}
\def\cis{CuIr$_2$S$_4$}

\def\cics{Cu(Ir$_{1-x}$Cr$_x$)$_2$S$_4$}
\def\range{$0 \le x \le 0.6$}
\def\ranget{$10~K \le T \le 300~K$}

\def\avo{AlV$_2$O$_4$}
\def\mto{MgTi$_2$O$_4$}
\def\irte2{IrTe$_2$}

\def\fd3m{Fd$\overline 3$m}
\def\p-1{P$\overline 1$}
\begin{document}

%
%
\title{{Cu(Ir$_{1-x}$Cr$_x$)$_2$S$_4$}: a model system for studying nanoscale phase coexistence at the metal-insulator transition}

\author{E.~S.~Bo\v{z}in,$^{1,*}$ K.~R.~Knox,$^{1}$ P.~Juh\'{a}s,$^{1}$ Y.~S. Hor,$^{2,\dag}$ J.~F.~Mitchell,$^2$ and S.~J.~L.~Billinge$^{1,3}$}

\affiliation{$^1$Condensed Matter Physics and Materials Science Department,
    Brookhaven National Laboratory, Upton, NY~11973}\email{bozin@bnl.gov}
\affiliation{$^2$Materials Science Division, Argonne
National Laboratory, Argonne, Illinois 60439 }
\altaffiliation[]{Current address: Department of Physics, Missouri University of Science and Technology, Rolla, MO 65409, USA}
\affiliation{$^3$Department of Applied Physics and Applied Mathematics, Columbia University, New York, NY~10027}

\date{\today}
%
%
%
\begin{abstract}
Increasingly, nanoscale phase coexistence and hidden broken symmetry states are being found in the vicinity of metal-insulator transitions (MIT), for example, in high temperature superconductors, heavy fermion and colossal magnetoresistive materials, but their importance and possible role in the MIT and related emergent behaviors is not understood.  Despite their ubiquity, they are hard to study because they produce weak diffuse signals in most measurements. Here we propose \cics\ as a model system, where robust local structural signals lead to key new insights.  We demonstrate a hitherto unobserved coexistence of a Ir$^{4+}$ charge-localized dimer phase and Cr-ferromagnetism. The resulting phase diagram that takes into account the short range dimer order, is highly reminiscent of a generic MIT phase diagram similar to the cuprates. We suggest that the presence of quenched strain from dopant ions  acts as an arbiter deciding between the competing ground states.
\end{abstract}

%
%

\maketitle

%
%

Metal-insulator transitions (MITs)~\cite{mott;bk74,zaane;prl85} remain one of the most fascinating phenomena in condensed matter physics~\cite{imada;rmp98,taguc;prb05}, especially in cases where emergent behavior appears close to the transition~\cite{georg;arocmp13} resulting in colossal effects, such as high-temperature superconductivity~\cite{bedno;zpb86,kamih;jacs06} and colossal magnetoresistance~\cite{jonke;p50}. For a remarkably broad range of materials, a similar phase diagram emerges with loss of a long-range ordered state crossing over to a homogeneous metallic state, but with a broad region where complex behavior, heterogenous states, difficult to detect broken local symmetries and nanoscale fluctuations become important~\cite{dagot;s05,dubi;n07,dagot;jpcm08}.  It is often precisely in this intermediate region where 
the interesting emergent behaviors occur. One of the major challenges is to study this complex region experimentally since the important physics seems to be closely related to details of the nature and interactions~\cite{georg;arocmp13} of the competing nano-scale heterogeneous phases, which cannot be easily studied using bulk average experimental probes.  New insights can be obtained from experimental probes that are spatially resolved with nanoscale resolution~\cite{pan;n01,kohsa;prl04,hanag;n04,kohsa;s07,kohsa;n08,lawle;n10,frati;n10,tao;prl09,tao;pnas11} and from probes that are sensitive to the local structure such as the atomic pair distribution function (PDF) analysis~\cite{bozin;prl00,jeong;prl05,qiu;prl05,bozin;prl07,malav;jpcm11} and extended x-ray absorption fine structure spectroscopy~\cite{booth;prl98,jiang;prb07,sunda;prl09,booth;prb11,haske;prl12}.  The important cuprates are especially challenging in this regard because of the small response of the lattice to the heterogenous electronic states.

\begin{figure}[tbf]
\includegraphics[width=0.45\textwidth,angle=0]{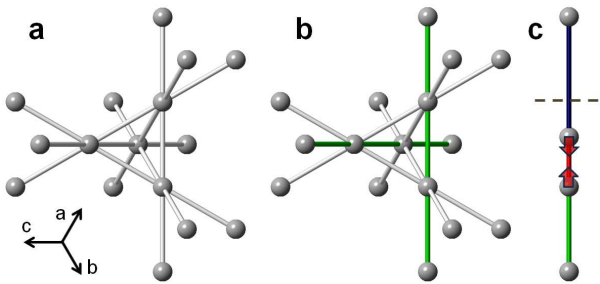}
\caption {{\bf Structural effect of dimerization in spinels.} (a) Regular pyrochlore sublattice in the high temperature cubic phase of transition metal spinels. All near neighbor transition metal distances (gray) are of equal average length ($t_{2g}$ are degenerate). The metal-metal bonds are shown, indicating degenerate interpenetrating 1D chains of ions (b) Structural distortion (for example tetragonal along $c$-axis) lifts the degeneracy (making the (110)-type directions (green) special). (c) Dimerization along the (110)-type chains occurring in the low temperature insulating phases results in a redistribution of bond-lengths: each dimer converts two average distances to a pair of short (red) and long (blue) distances. Details of the three dimensional ordering of the dimers then depend on the specifics of the transitional metal spinel family.}
\label{fig;dimerizationSketch}
\end{figure}

Among the wide variety of materials displaying MITs is a class of transition metal spinels with partially filled $t_{2g}$ orbitals~\cite{radae;njp05}. Exotic types of orbital and charge ordering, spin dimerization, and associated superstructures emerge in the low temperature insulating phases, such as octamers in \cis~\cite{radae;n02}, helices in \mto~\cite{schmi;prl04}, and heptamers in \avo~\cite{horib;prl06}. Appearance of these complex motifs is accompanied by a decrease in the magnetic susceptibility.  This behavior is understood first in \cis\ as the formation of Ir$^{4+}$-Ir$^{4+}$ structural dimers which results in the spins on the dimerized iridum ions forming a singlet (S=0) state. The interesting isomeric patterns that form in the different spinels depend on details of the chemistry and structure of the specific system and come about from different orderings of the dimers on the pyrochlore sublattice of corner shared tetrahedra of transition metal ions (Fig.~\ref{fig;dimerizationSketch}(a))~\cite{radae;n02}, but through a careful analysis of the local structure
and properties in \cis\ we show that the important physics driving the MIT is the charge localization and dimer formation itself rather than the dimer ordering and isomerization.

The structural dimerization is shown schematically in Fig.~\ref{fig;dimerizationSketch}.
%
%
In \cis\ the iridium ions are 3.5+ on average and sit on the corners of a 3D array of corner-shared tetrahedra, forming 3 degenerate sublattices, as
shown in Fig.~\ref{fig;dimerizationSketch}(a).  On cooling, a tetragonal distortion breaks the degeneracy of the three sublattices (Fig.~\ref{fig;dimerizationSketch}(b)).  One sublattice is then half filled and this band may lower its energy by forming a particularly short ``dimer" bond (Fig.~\ref{fig;dimerizationSketch}(c)). This is known in the undoped end-member because the dimers order over long-range giving rise to an easily measured structural phase transition, which correlates in temperature with changes in the conductivity and magnetic susceptibility. The state of ordered dimers that disappear on warming is analogous to the
undoped antiferromagnetic (AF) end-member in the cuprate high temperature superconductors in that a  long range ordered (LRO) broken symmetry insulating state is seen at low temperature.  It is thus interesting to see what happens in this system on doping.  This has been studied~\cite{endoh;prb03} in the case of Cr doping and a crossover to a metallic state, a MIT, where the metallic state has ferromagnetic LRO.

\begin{figure}[tbf]
\includegraphics[width=0.45\textwidth,angle=0]{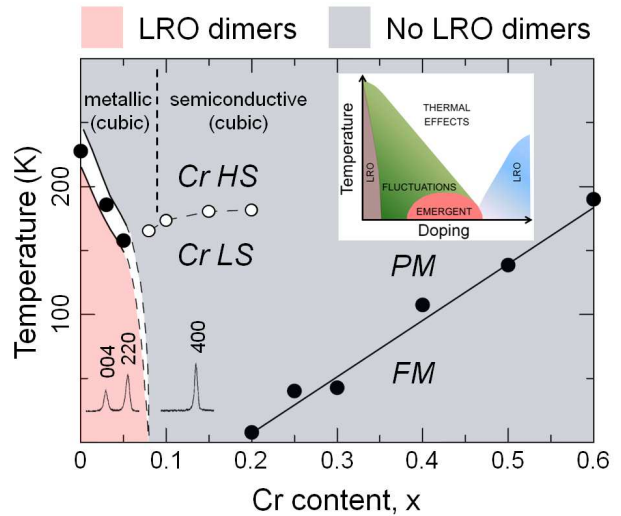}
\caption {{\bf \cics\ average phase diagram.} This is the canonical phase diagram largely reproduced from Endoh {\it et al.}~\cite{endoh;prb03}.  Regions with long range dimer order present (absent) are shaded in red (grey). A first order MIT is observed only close to the \cis\ end, with Cr-doping quickly suppressing the LRO. Insets at lower left show low temperature diffraction patterns for $x = 0$ (triclinic \p-1) and $x = 0.15$ (cubic \fd3m) that clearly show the change in crystallographic symmetry  going from the diamagnetic insulating (red) to the paramagnetic metallic, PM, (grey) phases. White symbols denote the onset of anomalies observed in resistivity and susceptibility for the intermediate $x$-range, originally attributed to Cr high spin (HS) to low spin (LS) transition~\cite{endoh;prb03}. For $x \ge 0.2$ the asymptotic ferromagnetic (FM) Weiss temperature increases linearly with increasing Cr-content (black circles). Inset at top right: a generic phase diagram discussed in the introduction.
}
\label{fig;averagePhaseDiagram}
\end{figure}

Although the MIT that takes place on warming in pure \cis\ is quite well understood~\cite{radae;n02,croft;prb03,khoms;prl05,radae;njp05,yagas;jpsj06,okamo;prl08,bozin;prl11}, the MIT on doping Cr~\cite{furub;jpsj94,endoh;prb03,nagat;chjp05} is not. The phase diagram as it is understood currently is shown in Fig.~\ref{fig;averagePhaseDiagram}~\cite{endoh;prb03}.  The long range dimer order is quickly suppressed with Cr doping and a metallic state emerges with ferromagnetic order appearing at higher
doping, presumably involving the Cr spins.  On the other hand, here we show that the {\it local} structure shows greater complexity with a crossover region in which a competition is taking place between competing ground-states leading to a nano-scale phase separation with a diamagnetic disordered-dimer phase coexisting with a ferromagnetically long-range ordered phase.
This is qualitatively similar to behavior in metal-insulator transitions in systems such as cuprate superconductors~\cite{tranq;n95,rezni;n06,xia;prl08,lawle;n10,mesar;s11,wu;n11,ghiri;s12,torch;nm13} and nickelates~\cite{tranq;prl93,tranq;prb95,wochn;prb98,li;prb03,hucke;prb04,hucke;prb06,abeyk;prl13} where hidden broken symmetry phases and nanoscale phase separation is seen in the vicinity of the MIT, though much easier
to study here because of the robust signal of the dimers in the local structure.  We believe that generic lessons learned in the \cis\ system can guide our understanding of MITs in these other systems since \cis\ exhibits a similar generic phase diagram, sketched in the inset to Fig.~\ref{fig;averagePhaseDiagram}.    The generic picture is that of a competition between two competing states, a paramagnetic
metallic state and a diamagnetic insulating charge-localized dimer state.  The  localized state is destabilized with respect to the delocalized metallic state on increased doping, but shows reentrant behavior that we explain due to dopant induced quenched disorder favoring the charge-localized dimer state. The material finally crosses over to a ferromagnetic metallic state at the Cr rich end of the phase diagram, but the metallic and insulating states coexist on the nanoscale over a wide range of doping and temperature, which can be well quantified in this system since the signal from the charge-localized dimer state has a distinctive local structural signature in the local structure.
%
%
%
%
%
\noindent{\bf Approach}
The local structure of the material is measured as a function of doping and temperature over a wide
range using the atomic pair distribution function analsyis (PDF) of X-ray diffraction data~\cite{egami;b;utbp13}.  High resolution PDFs, $G(r)$, where $r$ is the interatomic distance, were obtained by a Fourier transformation of powder diffraction data collected with short wavelength, high energy X-rays, according to $G(r) = {2\over\pi}\int_{Q_{min}}^{Q_{min}} Q[S(Q)-1]\sin Qr\>dQ$~\cite{farro;aca09}. The integrated data were corrected for experimental artifacts and normalized, to obtain the
total scattering structure function, $S(Q)$. This contains both Bragg and diffuse scattering and therefore information about atomic correlations on different length scales~\cite{qiu;prl05,bozin;prl07,bozin;s10}.  The data are analyzed by considering the evolution of features in the data-PDFs, in
particular peaks in the PDF that are sensitive to the presence or otherwise of the structural dimers, and by fitting structural models to the data.  This local structural information is then correlated with property measurements.

\begin{figure}[tbf]
\includegraphics[width=0.45\textwidth,angle=0]{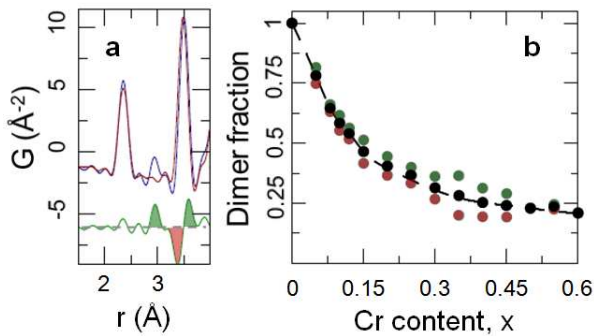}
\caption{\label{fig;dimerFraction} {\bf Dimer signal in the PDF and estimate of the fraction of dimerized Ir$^{4+}$.}
(a) Comparison of experimental PDFs at 300~K (red) and 10~K (blue) over a narrow $r$-range for \cis, with the difference curve (green) offset for clarity. Shaded features in the difference curve (color coded by green for dimer and red for loss peaks) are used in the dimer fraction evaluation. (b) Evolution of the dimer fraction with Cr-doping: the color code corresponds to that used for marking the features in the difference curve in (a), that were considered in the numeric integration analysis described in the Supplementary Material. Solid black symbols represent an arithmetic average of green and red values, while the dashed line is a guide to the eye.
}
\end{figure}

\noindent{\bf Results}\\
\noindent {\bf Evolution of dimers with doping.}
In the low temperature insulating phase of \cis\ pairs of Ir$^{4+}$ dimerize by moving closer together by a large 0.5~\AA~\cite{radae;n02} distance, resulting in the appearance of a distinct peak in the PDF at 3.0~\AA~\cite{bozin;prl11}. This is shown in Fig.~\ref{fig;dimerFraction}(a).
%
%
%
%
The low temperature PDF clearly displays an additional peak at around 3.0~\AA\ and this feature disappears in the high temperature data. A signature M-shape in the difference curve can also be observed, originating from the redistribution of PDF intensity from the position of the undistorted bonds before dimerization into short (dimerized) and long (non-dimerized) bonds, in accord with the dimerization invoked bond-length redistribution sketched in Fig.~\ref{fig;dimerizationSketch}(c).

\begin{figure}[tbf]
\includegraphics[width=0.425\textwidth,angle=0]{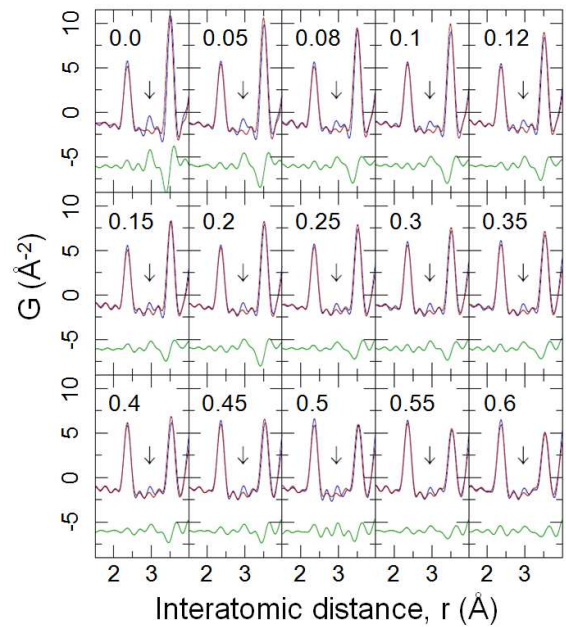}
\caption {{\bf Observation of doping dependence of the PDF dimer signature at low temperature.}  Comparison of experimental PDFs at 300~K (red) and 10~K (blue) over a narrow $r$-range for \cics (\range), with difference curve (green) offset for clarity. Each panel corresponds to a single composition $x$, as indicated. Arrows denote the position of the dimer PDF peak. The dimer signature clearly persists to $x=0.6$, though the signal is getting weaker.}
\label{fig;highTlowTPDFcomparison}
\end{figure}

This rather strong structural response allows the presence or absence of Ir$^{4+}$ dimers to be easily probed by a direct observation of this dimer peak~\cite{bozin;prl11}, and the M-shaped feature in difference curve. Dimers reveal themselves through this additional peak in the PDF irrespective of the length-scale of their ordering, allowing us to see the presence of localized charges even when they are only {\it local} in nature. For \cis\ this is demonstrated in the upper left panel of Fig.~\ref{fig;highTlowTPDFcomparison}~\cite{bozin;prl11}.
%
Although the low temperature long-range dimer order is swiftly suppressed by Cr-substitution resulting in an average cubic structure (Fig.~\ref{fig;averagePhaseDiagram}), the PDF analysis reveals that {\it on the nanoscale} dimers exist at low temperature within the entire composition range studied, since the dimer peak, indicated by an arrow in Fig.~\ref{fig;highTlowTPDFcomparison}, and the characteristic M-shaped difference curve, can be seen, albeit diminishing in amplitude with increasing doping, in all the low-$T$ data-sets.  Furthermore, careful quantitative analysis of the dimer signal in the difference curves of properly normalized PDFs enables an estimate of the dimer fraction to be carried out relative to the undoped endmember, as shown in Fig.~\ref{fig;dimerFraction}(b). The details of this estimation procedure are provided in the Supplementary Material. While the concentration of dimers decreases with doping, the distortion amplitude appears to be doping independent within the accuracy of our measurement. The PDF analysis presented here is based on total scattering data that do not discriminate between the elastic and inelastic scattering channels, and hence PDF does not distinguish whether the dimers are static or dynamic, though their obstervation in the local, but not average, structure establishes that they are not long-range ordered.\\

\begin{figure}[tb]
\includegraphics[width=0.45\textwidth,angle=0]{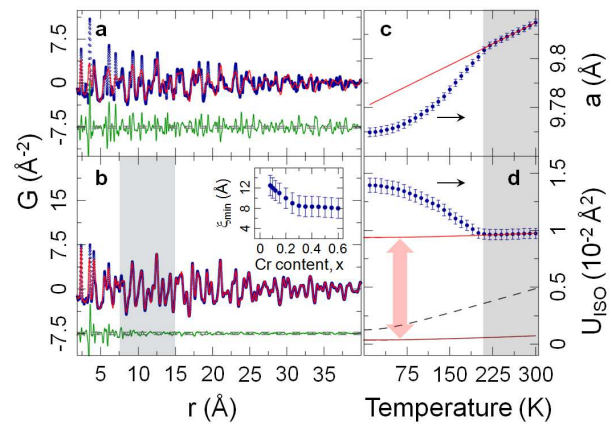}
\caption {{\bf Deviations from cubic \fd3m~ structure.} (a) \cis\ data at 10~K (open blue symbols), the best fit cubic model (solid red line), and the difference curve (green solid line) offset for clarity. Dashed lines are experimental uncertainties on the 2$\sigma$ level. (b) Same as (a) but for 40\% Cr sample. The gray area marks the $r$-region where the crossover from local to average behavior occurs. The temperature dependencies of the lattice parameter and Ir isotropic atomic displacement parameter (U$_{iso}$) for the 40\% Cr sample are shown in (c) and (d), respectively. Light red solid lines represent fits to the high temperature region of (c) (fit with a linear function) and (d) (fit with the Debye model). Deviations of the data from these trends are clearly observed at low temperature. (d) the Debye model with static offset (represented by the double arrow) set to zero (dark solid red line). The black dashed line represents the Debye model fit to Ir U$_{iso}$ for \cis\ data. An appreciably larger static offset for the 40\% Cr doped sample reflects the higher level of quenched disorder as compared to \cis. The inset to (b) shows the doping dependence of the lower $r$-boundary where deviations from the cubic model become apparent at low temperature.
}
\label{fig;deviationsFromCubic}
\end{figure}

\noindent {\bf Correlation length of dimer order.} Here we explore the correlation length of local dimer order across the phase diagram.
Since the PDF provides structural information on different lengthscales, this may be done by carrying out a variable $r$-range fit to the PDF of carefully chosen models~\cite{qiu;prl05}. Here we utilize the fact that the cubic spinel model, which describes the global structure well, fails for the dimer case. The observation of local dimers implies that the global cubic symmetry comes from an average over incoherent domains of locally ordered dimers where there is a lower symmetry within the domain.  We expect the cubic model to work well on lengthscales much larger than the local domain size, but to fail for short lengthscales dominated by the intra-domain signal.   In Fig.~\ref{fig;deviationsFromCubic}(a) and (b) fits of the cubic model to the low temperature data of 0\% Cr (exhibiting long range ordered dimers) and 40\% Cr (no long range ordered dimers, dimer signal relatively weak) samples are shown, respectively. The model fails for \cis\ on all lengthscales, as expected. On the other hand, for the 40\% Cr sample with \fd3m global symmetry the cubic model fails only on lengthscales shorter than $\sim 1$~nm, as evident from the difference curve extending out of the 2$\sigma$ uncertainty parapet in Fig.~\ref{fig;deviationsFromCubic}(b).
The correlation length determined in this way for all the samples is plotted in the inset to Fig.~\ref{fig;deviationsFromCubic}(b), where the error bars indicate a range of uncertainty associated with determining this crossover marked as the grey band in Fig.~\ref{fig;deviationsFromCubic}(b), with further discussion of how this was obtained provided in the Supplementary Material.
No significant temperature variation of this bound could be established within the accuracy of our measurement. Although our analysis does not explicitly address the nature of the dimer order, it is interesting to contemplate what such short range dimer order may include. While the observed correlation length bound extends only to near neighbor dimers along the (110)-chains, it encompasses both near neighbor and next near neighbor dimers spanning the octamers, suggesting a three dimensional character of the dimer correlations.\\

%
%

%
\begin{figure}[tb]
\includegraphics[width=0.475\textwidth,angle=0]{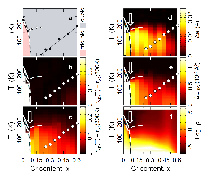}
\caption {{\bf Reassessment of the ($x,T$) phase diagram of \cics.} (a) Average \cics\ phase diagram shown in Fig.~\ref{fig;averagePhaseDiagram}~\cite{endoh;prb03}. (b) ($x,T$) dependence of the cubic model fit residual, R$_{wp}$, for the \fd3m\ cubic model fit over the high-$r$ 15-40\AA\ PDF range, reproducing the behavior in (a). (c) Same as (b) but for the PDF refinements including the low-$r$ region, revealing the local dimers in the R$_w$ parameter. (d) ($x,T$) dependence of the differential \fd3m\ lattice parameter. (e) ($x,T$) dependence of the differential Ir $U_{iso}$. See text for definitions. (f) ($x,T$) dependence of $Log~\rho$, after reference~\cite{endoh;prb03}. In (b)-(e): Weiss temperature of our samples is shown as white solid symbols. Dashed lines reproduce phase lines discussed in Fig.~\ref{fig;averagePhaseDiagram}. Vertical arrow denotes a feature discussed in the text.
}
\label{fig;dimerPhaseDiagram}
\end{figure}
%

\noindent {\bf Evolution of dimers with temperature.} Next we address the temperature dependence of the dimer existence. This may be accomplished by plotting relevant structural parameters, such as lattice parameter, atomic displacement parameters (ADPs, $U_{iso}$), and cubic model fit residual, R$_{wp}$, obtained from average structure fits to the PDF data.
While the average crystallographic model does not see the broken symmetry state directly, anomalous behavior of these parameters is to be expected as the nanoscale state sets in, and can be a sensitive indirect confirmation of their presence~\cite{abeyk;prl13}, for example, R$_{wp}$ will increase when the local structure is not well explained by the cubic model. The temperature evolution of the cubic lattice parameter and Ir $U_{iso}$ from the 40\% Cr sample on cooling are shown in Fig.~\ref{fig;deviationsFromCubic}(c) and (d), respectively. On cooling down from 300~K (gray region) both parameters evolve smoothly displaying canonical behavior, as evident from the linear fit to the lattice parameter, and the Debye model fit to the ADP of Ir (see Supplementary Material). However, dramatic deviations from these trends are observed at $\sim$180~K, with the lattice parameter decreasing rapidly and $U_{iso}$ of Ir exhibiting an upturn. These deviations mark the onset temperature of the local dimer formation on cooling.

Using this same approach it is then possible to determine the evolution of the dimer formation across the whole phase diagram.
In Fig.~\ref{fig;dimerPhaseDiagram} we plot the evolution of different system parameters vs. doping and temperature.  For example, the existence of a global cubic structure may be found by excluding the low-$r$ region of the PDF from fits of the cubic model.  This is shown in Fig.~\ref{fig;dimerPhaseDiagram}(b) and is in good agreement with previous crystallographic results displayed in Fig.~\ref{fig;dimerPhaseDiagram}(a)~\cite{endoh;prb03}. However, if we  include the low-$r$ region of the PDF in the fit a very different picture emerges (Fig.~\ref{fig;dimerPhaseDiagram}(c)): the {\it nanoscale} phase diagram of \cics\ that maps out the ($x$,$T$) evolution of local dimers. A similar picture emerges from other measures of the local dimers such as the differential lattice parameter, $\Delta a$, and differential Ir isotropic atomic displacement parameter, $\Delta$U$_{iso}$ (Fig.~\ref{fig;dimerPhaseDiagram}(d) and (e)).  These quantities are obtained by subtracting the observed behavior of the parameters from the expected behavior in the absence of local dimer formation, as determined by an extrapolation to low temperature of the high-temperature behavior using standard assumptions such as as Debye behavior for the ADPs.  All these measures reveal a previously undetected dome of local dimers appearing, roughly peaked at $\sim 200$~K at around $x=0.3$.  These strategies for searching for the presence
of local, broken symmetry, states carry over, in principle, to other material systems of interest such as the cuprates and nickelates.\\
%
\begin{figure}[tb]
\includegraphics[width=0.45\textwidth,angle=0]{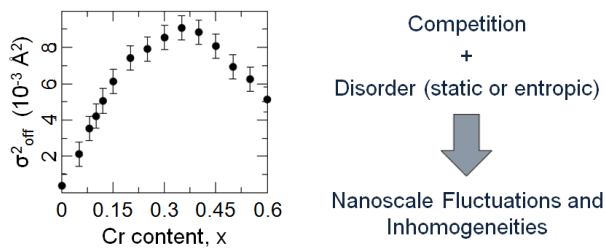}
\caption {{\bf Measure of quenched disorder in \cics.} The Debye model offset parameter, a sensitive measure of quenched disorder, versus Cr-doping, as obtained from the fits of this model to the Ir $U_{iso}(T)$ (left). Schematic of ingredients leading to nanoscale fluctuations and inhomogeneities (right).
}
\label{fig;strainMeasures}
\end{figure}
%

%
%

\noindent {\bf Discussion}\\
The new information provided by this study leads to a completely different interpretation to the standard one~\cite{endoh;prb03} of the phase line close to 180~K in the intermediate doping regime, marked by white symbols in Figs.~\ref{fig;averagePhaseDiagram} and ~\ref{fig;dimerPhaseDiagram}(a), and by dashed white lines in Figs.~\ref{fig;dimerPhaseDiagram}(b)-(f).  This was originally interpreted from analysis of magnetic susceptibility data as coming from a low spin (LS, $s = 1/2$) to high spin (HS, $s = 3/2$) crossover of the doped Cr ions~\cite{endoh;prb03}.
It is clear from this work that the anomalous loss in magnetic susceptibility in doped samples on cooling is actually due to the formation of local dimers that are not apparent in the average structure.\\

It is interesting to ask whether it is the global or local structure which is more important for determining the transport properties of the material.  To explore this, we plot the resistivity vs $x$ and $T$ in Fig.~\ref{fig;dimerPhaseDiagram}(f).  The resemblance of this to the local phase diagram suggests that the local structure is very important for determining (and understanding) the transport properties. On warming the dimer state is replaced by the charge delocalized paramagnetic metallic state.  However, on doping at low temperature, as we show here, it is replaced by a disordered or short-range ordered dimer phase.

One of the most interesting parts of the  phase diagram occurs around $x=0.07$ and $T\sim 175$~K (marked by an arrow in Fig.~\ref{fig;dimerPhaseDiagram} (c)-(f)) where local dimers disappear on doping, but are re-entrant at higher doping.
The doping quickly destabilizes the ordered dimer phase yielding the paramagnetic metallic phase, and indeed the dimers are destroyed even in the local structure, but on further doping the localized, but now disordered, dimers are again preferred.
There are elastic strain costs associated with localizing charges as dimers when they are not ordered over long range.

It may be possible to explain the reentrant behavior if local lattice strain fields around doped Cr ions may be utilized by dimers to form without paying an additional strain energy cost. It is not known if the doped Cr is in the 3+ or 4+ state because copper can also alter its valence to maintain charge neutrality.  However, either way the doped Cr is smaller (0.615~\AA\ and 0.55~\AA\ for Cr$^{3+}$ or as Cr$^{4+}$, respectively) than the Ir that it replaces (0.68~\AA\ and 0.625~\AA\ for Ir$^{3+}$ or as Ir$^{4+}$, respectively). This is in line with the observation in \cics\ that the average lattice parameter decreases on doping with Cr~\cite{endoh;prb03}.  The doped Cr ions therefore result in a compressive strain of the lattice in their vicinity.    The dimer formation also results in a decrease in the unit cell volume~\cite{furub;jpsj94,radae;n02}, and also pressure stabilizes the charge-localized dimer state~\cite{oomi;jmmm95,garg;ssc07}. It is thus plausible that dimers will form preferrentially in the compressive strain field of the quenched defects.

As the phase diagram is traversed at 175~K, at a doping high enough to destroy dimer long range order, but still low, there are many dimers to be accommodated but few Cr ions and the paramagnetic metallic phase becomes energetically preferred.  However, with increased Cr doping the increase in the number of Cr sites and the decrease in the number of dimers to be accommodated presumably are better matched, and the the charge-localized dimer state may be again stabilized locally, though without the dimers ordering over long range.  This would explain the reentrant and dome-like behavior of the local dimer state that we observe and is also consistent with the doping dependence of the static disorder as measured from the offset parameter in the Debye model fits to the Ir $U_{iso}(T)$ shown in the left panel of Fig.~\ref{fig;strainMeasures} (see Supplementary Material for details).  The electrical properties are sensitive to the local dimer state and the conductivity of the sample results in a similar dome-like behavior to the local-dimer state (Fig.~\ref{fig;dimerPhaseDiagram}(f)), reinforcing the importance of characterizing the local nature of the material to understand the properties.  This strain-biasing effect can be expected to be broadly applicable to a wide range of systems exhibiting doping induced MITs and generically shows how the competition between ground-states, in the presence of quenched disorder, can lead to a nanoscale coexistence of both of the competing phases (Fig.~\ref{fig;strainMeasures}).  It just requires that the competing phases have slightly different average bond lengths and the dopant ions introduce local strains into the lattice~\cite{billi;prb02}.

A ferromagnetic metallic state appears in the region of high Cr doping region~\cite{endoh;prb03}, introducing a third competing state into the picture.  Our PDF measurements cannot distinguish the paramagnetic metallic phase seen at low doping and the ferromagnetic metallic state seen at higher doping, but the ferromagnetic signal is also detected in our magnetization measurements for as low as $x=0.25$, implying that there is a wide region of coexistence of the charge-localized dimer and ferromagnetic states at low temperature, which can only be explained as a nanoscale coexistence of insulating charge-localized dimer and metallic ferromagnetic phases.
%
\begin{figure}[tb]
\includegraphics[width=0.45\textwidth,angle=0]{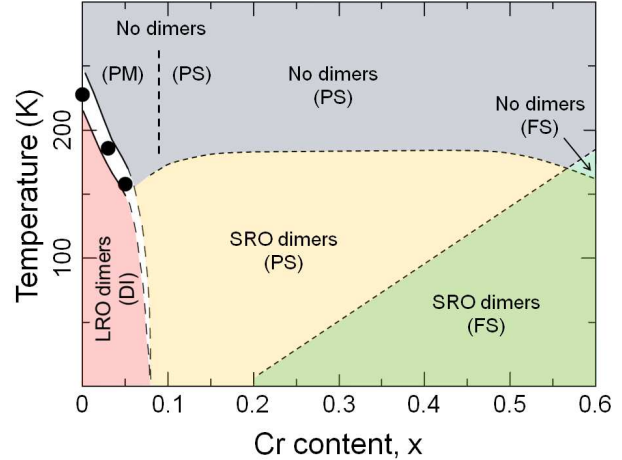}
\caption {{\bf Revised phase diagram for \cics .} Regions with dimers being long range ordered (LRO), short range ordered (SRO), and absent are identified. Dominant magnetic and transport characteristics are also noted, with D=diamagnetic, P=paramagnetic, F=ferromagnetic, M=metallic, I=insulating, and S=semiconducting. The average structure is triclinic in the diamagnetic insulating phase, and cubic elsewhere. The local structure is distorted on a nanometer lengthscale in regions where SRO dimers exist.
}
\label{fig;revisedPhaseDiagram}
\end{figure}
%

In summary, this study provides a revised \cics\ phase diagram, Fig.~\ref{fig;revisedPhaseDiagram}, and shows how enormous insight into the physics of doped systems with MITs can be obtained from a systematic study of local structure over wide ranges of temperature and doping. \cics\ can serve as a model system for this because of the robustness of the structural signal associated with the charge-localized dimer state.
The general notion that broken symmetry states can persist, undetected, in probes of average structure over wide ranges of doping and temperature and may explain anomalies in bulk properties such as transport and magnetization, is likely to be carried over to other doped systems at the metal-insulator boundary.\\

\noindent {\bf Methods}\\
\noindent {\bf Sample synthesis and characterization.} Polycrystalline samples of \cics\ with \range\ were prepared by a standard solid state route in sealed, evacuated quartz ampoules. Stoichiometric quantities of the metals and elemental sulfur were thoroughly mixed, pelletized, and sealed under vacuum.
The ampoules were slowly heated to 650-750~$^{o}$C and held at this temperature for several weeks with intermediate grinding
and pressing.  The products were found to be single phase based on laboratory x-ray powder diffraction. DC susceptibility data were measured on cooling in a 1~T field using a Quantum Design PPMS.  Resistivity was measured using a standard four-terminal technique.\\

\noindent {\bf Total scattering experiments.}
PDF data for \ranget\ were obtained using standard protocols~\cite{egami;b;utbp13} from synchrotron x-ray total scattering experiments carried out at the 11-IDC beamline of the Advanced Photon Source at Argonne National Laboratory. The setup utilized a 114.82~keV x-ray beam ($\lambda = 0.108$~\AA) and a Perkin Elmer amorphous silicon detector, and a closed cycle helium refrigerator. The raw 2D data were integrated and converted to intensity versus $Q$ using the software Fit2D~\cite{hamme;hpr96}, where $Q$ is the magnitude of the scattering vector.
Data reduction and Sine Fourier transform of measured total scattering structure functions up to a high momentum transfer of $Q_{max}= 27$~\RAA\ was carried out using the {\sc PDFgetX3}~\cite{juhas;jac13} program. PDF structure refinements were carried out using the {\sc PDFgui} program suite~\cite{farro;jpcm07}.\\

\noindent {\bf References}

%
%

%
%
\noindent {\bf Acknowledgements}
Work at Brookhaven National Laboratory is supported by the U.S. Department of Energy
Office of Science (DOE-OS) under Contract no. DE-AC02-98CH10886. Work at Argonne National
Laboratory, which is operated by UChicago Argonne LLC, and the Advanced Photon Source facility,
are supported under the U.S. DOE-OS Contract No. DE-AC02-06CH11357.

%
%
\noindent {\bf Author contributions}
E.S.B. and S.J.L.B. designed this work; Y.S.H. and J.F.M. synthesized the samples and carried out magnetic characterizations; E.S.B. and P.J. carried out total scattering experiments and PDF data reduction; E.S.B., P.J., and K.R.K. performed PDF analysis; E.S.B. and S.J.B. wrote the paper; all the authors participated in discussions of the results as well as preparing the manuscript.

%
%

\noindent {\bf Additional information}\\
\noindent {\bf Supplementary Information} accompanies this paper.

\noindent {\bf Competing financial interests:} The authors declare no competing financial interests.

\end{document}